\documentclass[
 reprint,
superscriptaddress,
 amsmath,amssymb,
prl,
floatfix,
]{revtex4-2}

\usepackage{graphicx}
\usepackage{dcolumn}
\usepackage{bm}
\usepackage[utf8]{inputenc}
\usepackage[T1]{fontenc}
\usepackage[colorlinks,urlcolor=blue,citecolor=blue,linkcolor=blue]{hyperref}
\usepackage{textcomp}

\begin{document}

\preprint{VCSEL-BEC}
\title{Bose-Einstein condensation of photons in a vertical-cavity surface-emitting laser}

\author{M. Pieczarka}
 \email{maciej.pieczarka@pwr.edu.pl}

 \affiliation{Department of Experimental Physics, Faculty of Fundamental Problems of Technology, Wroc\l{}aw University of Science and Technology, Wyb.~Wyspia\'nskiego 27, 50-370 Wroc\l{}aw, Poland}
 
\author{M. G\k{e}bski}%
 \affiliation{Institute of Physics, Lodz University of Technology, ul. W\'{o}lcza\'{n}ska 217/221, 93-005 \L\'{o}d\'{z}, Poland}
 
\author{A. N. Piasecka}

 \affiliation{Department of Experimental Physics, Faculty of Fundamental Problems of Technology, Wroc\l{}aw University of Science and Technology, Wyb.~Wyspia\'nskiego 27, 50-370 Wroc\l{}aw, Poland}
 
\author{J. A.~Lott}

 \affiliation{Institute of Solid State Physics and Center of Nanophotonics, Technical University Berlin,
Hardenbergstra\ss e 36, 10623 Berlin, Germany}

\author{A. Pelster}%
 \affiliation{Department of Physics and Research Center OPTIMAS, Rheinland-Pf\"{a}lzische Technische Universit\"{a}t Kaiserslautern-Landau,
Erwin Schr\"{o}dinger Stra\ss e 46, 67663 Kaiserslautern, Germany}

\author{M. Wasiak}%
 \affiliation{Institute of Physics, Lodz University of Technology, ul. W\'{o}lcza\'{n}ska 217/221, 93-005 \L\'{o}d\'{z}, Poland}

 \author{T. Czyszanowski}%
 \affiliation{Institute of Physics, Lodz University of Technology, ul. W\'{o}lcza\'{n}ska 217/221, 93-005 \L\'{o}d\'{z}, Poland}

\maketitle

\noindent \textbf{Many bosons can occupy a single quantum state without a limit. This state is described by quantum-mechanical Bose-Einstein statistics, which allows the formation of a Bose-Einstein condensate at low temperatures and high particle densities. Photons, historically the first considered bosonic gas, were late to show this phenomenon, which was observed in rhodamine-filled microlaser cavities and doped fiber cavities. These more recent findings have raised the natural question as to whether condensation is common in laser systems, with potential technological applications. Here, we show the Bose-Einstein condensation of photons in a broad-area vertical-cavity surface-emitting laser with positive cavity mode-gain peak energy detuning. We observed a Bose-Einstein condensate in the fundamental transversal optical mode at the critical phase-space density. The experimental results follow the equation of state for a two-dimensional gas of bosons in thermal equilibrium, although the extracted spectral temperatures were lower than those of the device. This is interpreted as originating from the driven-dissipative nature of the device and the stimulated cooling effect. In contrast, non-equilibrium lasing action is observed in the higher-order modes in a negatively detuned device. Our work opens the way for the potential exploration of superfluid physics of interacting photons mediated by semiconductor optical non-linearities. It also shows great promise for enabling single-mode high-power emission from a large aperture device.}

At the beginning of the 20th century, Albert Einstein extended the statistical theory of Satyendra Nath Bose to describe massive particles and made the pioneering prediction of the Bose-Einstein condensate (BEC) below a critical temperature~\cite{Einstein1925}. BEC is characterized by both saturation of occupation in the excited states and condensation in the ground energy state of the system~\cite{Pitaevskii2016}. Seventy years after its theoretical prediction, this macroscopic quantum phenomenon was first observed directly in dilute clouds of atomic gases at temperatures close to absolute zero~\cite{Anderson1995,Davis1995}. The reason for such a low critical temperature is that it is inversely proportional to the mass of the boson. Therefore, a heavy particle gas must be extremely cold to reach the transition point. However, if we consider the mass as the parameter of energy dispersion, then we can find a bosonic quasiparticle described with a dispersion of large curvature, and hence with a quite small effective mass, which enables condensation at elevated temperatures. This concept has been realized in a variety of bosonic quasiparticle systems, such as magnons~\cite{demokritov2006}, excitons~\cite{Butov2002,Wang2019,Alloing2014}, and plasmons~\cite{Hakala2018}, as well as hybrid excitations of strongly coupled systems of exciton and photons, namely cavity polaritons~\cite{Kasprzak2006,Sun2017}.
\begin{figure*}[ht]
\centering
\includegraphics[width=\textwidth]{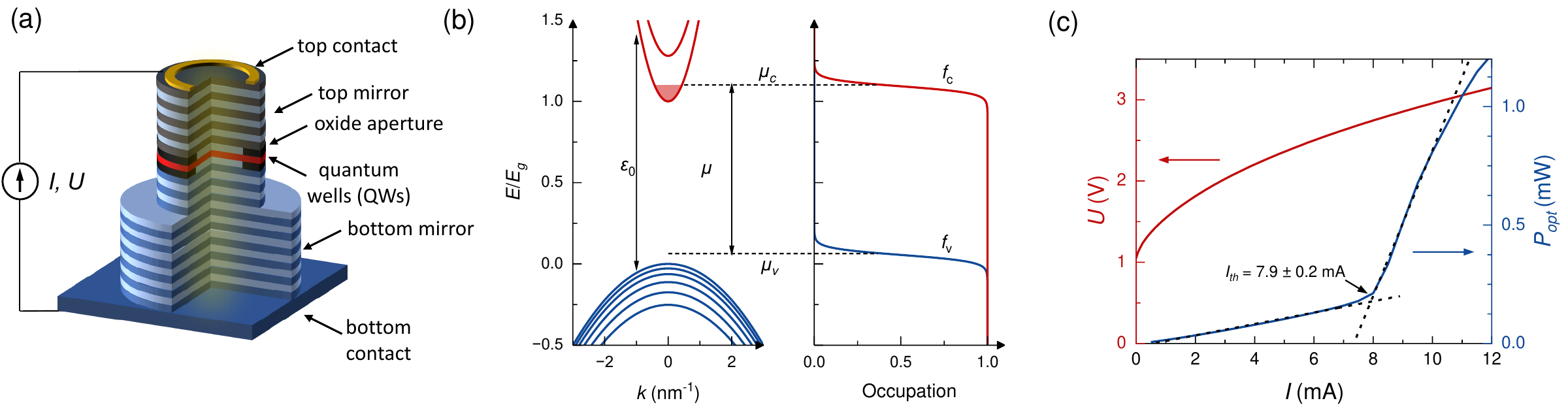}
\caption{\textbf{Basic properties of a VCSEL.} \textbf{(a)} Scheme of the investigated VCSEL devices with all main components indicated by arrows. \textbf{(b)} Simplified picture of the conduction and valence subbands confined in the QWs expressed in the in-plane wavevector (left). The occupations of the conduction band $f_c$ and the valence band states $f_v$ expressed with Fermi-Dirac distributions of different quasi-Fermi levels $\mu_c$ and $\mu_v$, respectively. 
$\varepsilon_0$ is the energy of the fundamental cavity mode, which is larger than the semiconductor band gap. \textbf{(c)} Output power-current-voltage (LIV) characteristics of the BEC device.}
\label{fig:Fig1}
\end{figure*}

Photons, on the other hand, have been out of the picture for many years because they represent a massless gas with linear energy dispersion and a trivial, null ground state. In principle, the number of particles is not conserved, i.e.~in a blackbody radiation model in thermal equilibrium the chemical potential vanishes, and therefore condensation cannot occur. Nevertheless, over years of research many analogies have been drawn between laser physics and atomic BEC physics, yielding a more detailed understanding of these two worlds. Eventually, a system that meets all the requirements of an equilibrium photon BEC was obtained in a laboratory tabletop system of a microcavity filled with a rhodamine solution~\cite{Klaers2010}. Remarkably, this system clearly demonstrated many textbook effects of a non-interacting condensate of bosons, from thermodynamic and caloric properties~\cite{Damm2016,Damm2017} to quantum-statistical effects~\cite{Schmitt2014,Walker2020}. Moreover, the driven-dissipative nature of this system beyond equilibrium has been demonstrated~\cite{Walker2018}, and the phase boundaries between photon BECs and non-equilibrium lasing have been investigated extensively~\cite{Kirton-Keeling, Radonjic2018}. However, rhodamine-based photon BECs are limited by their weak and slow thermo-optical nonlinearity \cite{Klaers2011}, which has so far prevented the observation of static or dynamic superfluid effects. Pioneering observations have stimulated the search for BEC conditions in other laser systems, such as fiber cavities~\cite{Weill2019} and semiconductor lasers~\cite{Barland2021, Kammann2012, Bajoni2007}, to enable true technological applications outside of the laboratory environment and to find a material system with non-negligible and fast non-linearities. 

Here, we demonstrate a photon BEC in a well-established semiconductor device, a large aperture electrically driven vertical-cavity surface-emitting laser (VCSEL) at room temperature. By testing devices with different energy detunings between the cavity fundamental mode $\varepsilon_0$ and the quantum well (QW) fundamental transition $\varepsilon_{\rm QW}$, defined as $\Delta = \varepsilon_0-\varepsilon_{\rm QW}$, we found a homogeneous BEC of photons with a thermalized spectrum. This occurred for both $\Delta>0$ and standard non-equilibrium laser operation at higher-order modes in another device of the same geometry but with $\Delta<0$. In the BEC regime, we found that the photonic gas thermalizes to temperatures below the temperature of the VCSEL, suggesting that it is not fully equilibrated with the optical gain medium. Nevertheless, the extracted temperatures, chemical potentials, and photon densities allowed us to experimentally determine the equation of state (EOS), which follows the behavior of a 2D Bose gas in thermal equilibrium. 

The device under study is an epitaxially grown oxide-confined VCSEL with a large 23\,\textmu m-diameter aperture, emitting around 980\,nm. The VCSEL is designed for simultaneous high bandwidth, high optical output power, and moderate to high wall plug efficiency~\cite{Haghighi2019, Haghighi2019_2} (see Methods for details). We drive our semiconductor device at room temperature with direct current, by applying a constant voltage across the laser diode (see Fig.~\ref{fig:Fig1}(a)). This sets the non-equilibrium distribution of carriers in the QW region, as the separation of the quasi-Fermi levels for electrons in the conduction band states $\mu_c$ and holes in the valence band states $\mu_v$ is proportional to the applied voltage. Due to the sub-picosecond relaxation of carriers within the bands~\cite{Knox1986}, the electrons and holes are in equilibrium with the device. Hence, both gases can be described with separate Fermi distributions, with different quasi-Fermi levels setting the occupation in both bands (see Fig.~\ref{fig:Fig1}(b)~\cite{Coldren2012}). 
\begin{figure*}[ht]
\centering
\includegraphics[width=\textwidth]{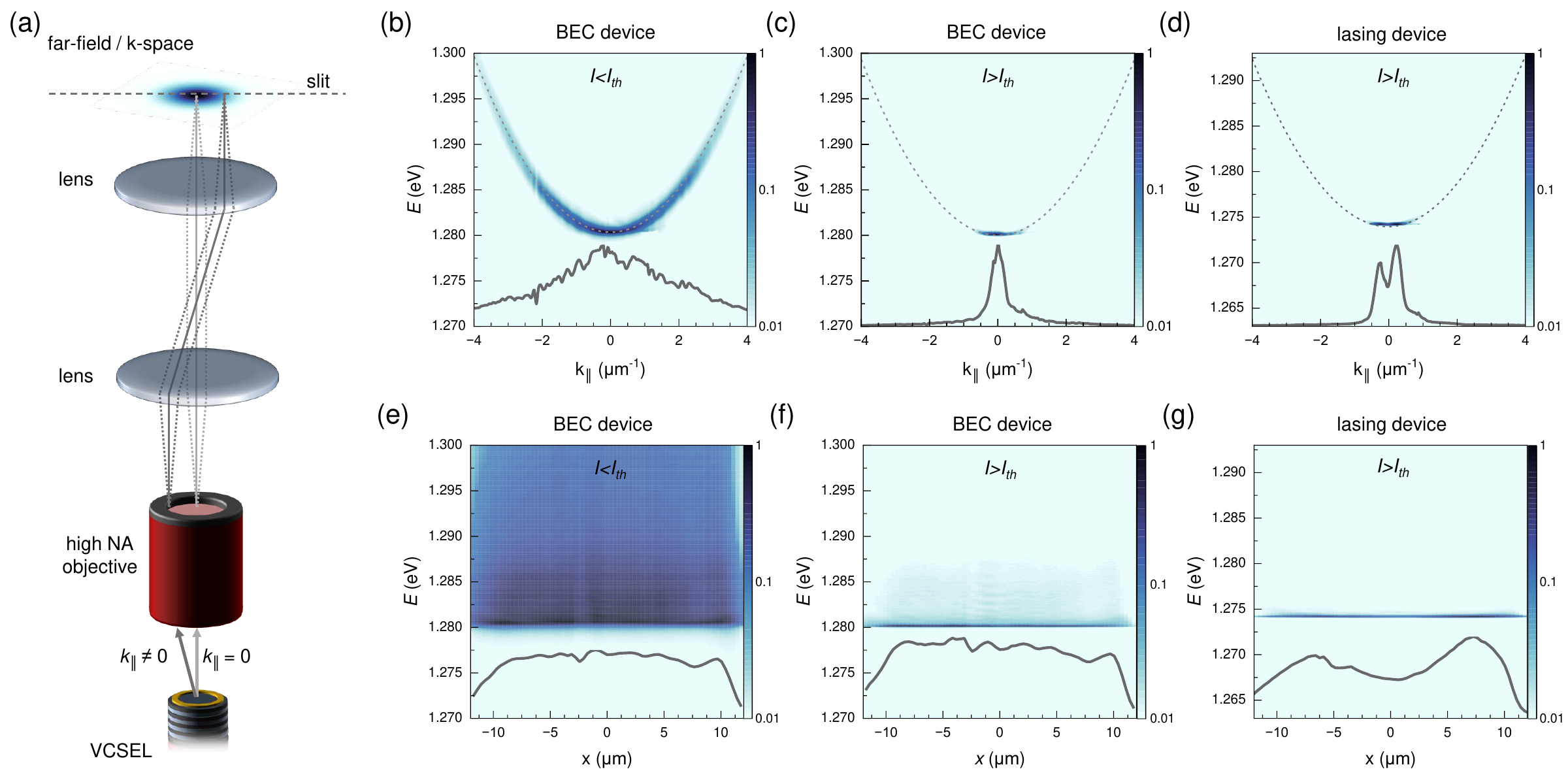}
\caption{\textbf{Momentum-space and real-space spectra of the BEC and a laser device}. \textbf{(a)} Scheme of the experimental setup used for momentum-space imaging. The back focal plane of the microscope objective is imaged onto the entrance slit of the monochromator, and then it is dispersed to the CCD camera, enabling probing of the spectral information at the center cut of the momentum space. \textbf{(b)} Momentum-space spectrum of the BEC device below ($I = 5$\,mA) and \textbf{(c)} above ($I=8.5$\,mA) the condensation threshold showing the narrowing from thermal distribution to the ground state $k_{\parallel} \approx 0$. \textbf{(d)} Momentum-space spectrum of the lasing device in the higher-order mode above the lasing threshold ($I=6.3$\,mA). \textbf{(e)},\textbf{(f)} Real-space spectra of the BEC device below and above the threshold showing homogeneity of the gas. \textbf{(g)} Real-space spectrum of the lasing device that presents the domination of the higher-order mode. All color scales are logarithmic to enhnace the visibility of high-energy states. Insets in \textbf{(b-g)} represent the normalized energy-integrated spectra in linear scale. }
\label{fig:Fig2}
\end{figure*}

Let us assess the essential conditions for obtaining a photon BEC in a VCSEL. In electrically driven semiconductors, excited electrons and holes can recombine, emitting photons. Thus, the condition of chemical equilibrium can be established if the chemical potential of photons is equal to $\mu = \mu_c-\mu_v$, in close analogy to a photochemical reaction~\cite{Wurfel1982}. This well-defined chemical potential of a photon gas is essential for obtaining a BEC at equilibrium. Another key ingredient is the detailed balance condition between emission and absorption, which was explored in the first demonstrations of photon BECs based on organic laser dyes~\cite{Klaers2010}. This condition is also met in semiconductors, where the ratio between emission and absorption rates is expressed as the van Roosbroeck-Schockley relation $R_{\rm abs}(\varepsilon)/R_{\rm em}(\varepsilon)=\exp(\frac{\varepsilon-\mu}{k_{B}T})$ (see Methods)~\cite{Coldren2012,Bhattacharya2012}. Hence, the thermalization of light occurs after a few cycles of spontaneous emission and absorption events before the photons escape the cavity through the mirror. Such energy exchange with the active medium enables the photon gas to establish both a chemical potential and a temperature. Eventually, it leads to a modified Bose-Einstein (BE) distribution of photons, which can be derived from the laser rate equations (see Methods):
\begin{equation}
N(\varepsilon) = \frac{1}{\exp(\frac{\varepsilon-\mu}{k_{B}T})-1+\Gamma(\varepsilon)}\, .
    \label{eq:eq1}
\end{equation} 
Here, the correction parameter $\Gamma(\varepsilon)=\gamma(\varepsilon)/R_{\rm em}(\varepsilon)$ represents the ratio of the photon decay rate from the passive cavity $\gamma$ and the spontaneous emission rate $R_{\rm em}$ to the photon mode at a given energy $\varepsilon$. Consequently, this correction parameter can be treated as a measure of the  degree of thermalization. It is expected to be small if many photon emission-absorption cycles occur before the photons escape the cavity. At this limit, Equation~(\ref{eq:eq1}) approaches the Bose-Einstein distribution. Based on our numerical modeling and experimental measurements, we estimated this ratio for the fundamental mode at $\Gamma(\varepsilon_0) \approx 0.008$ (see Methods and Supplementary Information for details), ensuring that we obtained a thermalized photon gas in our system.

According to standard semiconductor laser theory, the Bernard-Duraffourg condition~\cite{Bernard1961}, which is essential for non-equilibrium lasing, is met when the value of the chemical potential exceeds the energy of an optical mode $\mu>\varepsilon$. This creates a positive optical gain at this energy~\cite{Coldren2012}, so thermalization is expected to dominate below this limit. Therefore, we probed devices with different cavity-QW gain detunings $\Delta$. We used the side effect of epitaxial growth that the resulting layers are not homogeneous throughout the entire wafer and have a tendency to become thinner towards the edge~\cite{Mogg2004,Balili2006,Hegarty1999}. Phenomenologically, this affects the cavity energy shifts more than the spectral shifts of the gain. Thus, close to the center of the three-inch wafer we probe the device with $\Delta <0$, which is the standard designed detuning for high-power and high-temperature lasing operation, which we denote as the lasing device. In contrast, close to the wafer edge the detuning becomes positive, and the device is expected to operate in the thermalized BEC regime, which we denote as the BEC device. Although we cannot directly measure the precise value of the detuning, we observed a stark contrast in the performance of devices from these distant positions on the sample, supporting our assumptions.  

The electrical and total output power characteristics on the driving current of the BEC device are shown in Fig.~\ref{fig:Fig1}(c), and the results of the lasing device are summarized in the Supplementary Information. The data show all the standard features of a laser, the electrical characteristics of a diode, and the emission threshold current $I_{\rm th}$. However, the device is characterized by significant spontaneous emission below $I_{\rm th}$. Therefore, the information contained in the spectral characteristics of the device must be examined to distinguish between a BEC and a lasing state.
\begin{figure*}[ht]
\centering
\includegraphics[width=.75\textwidth]{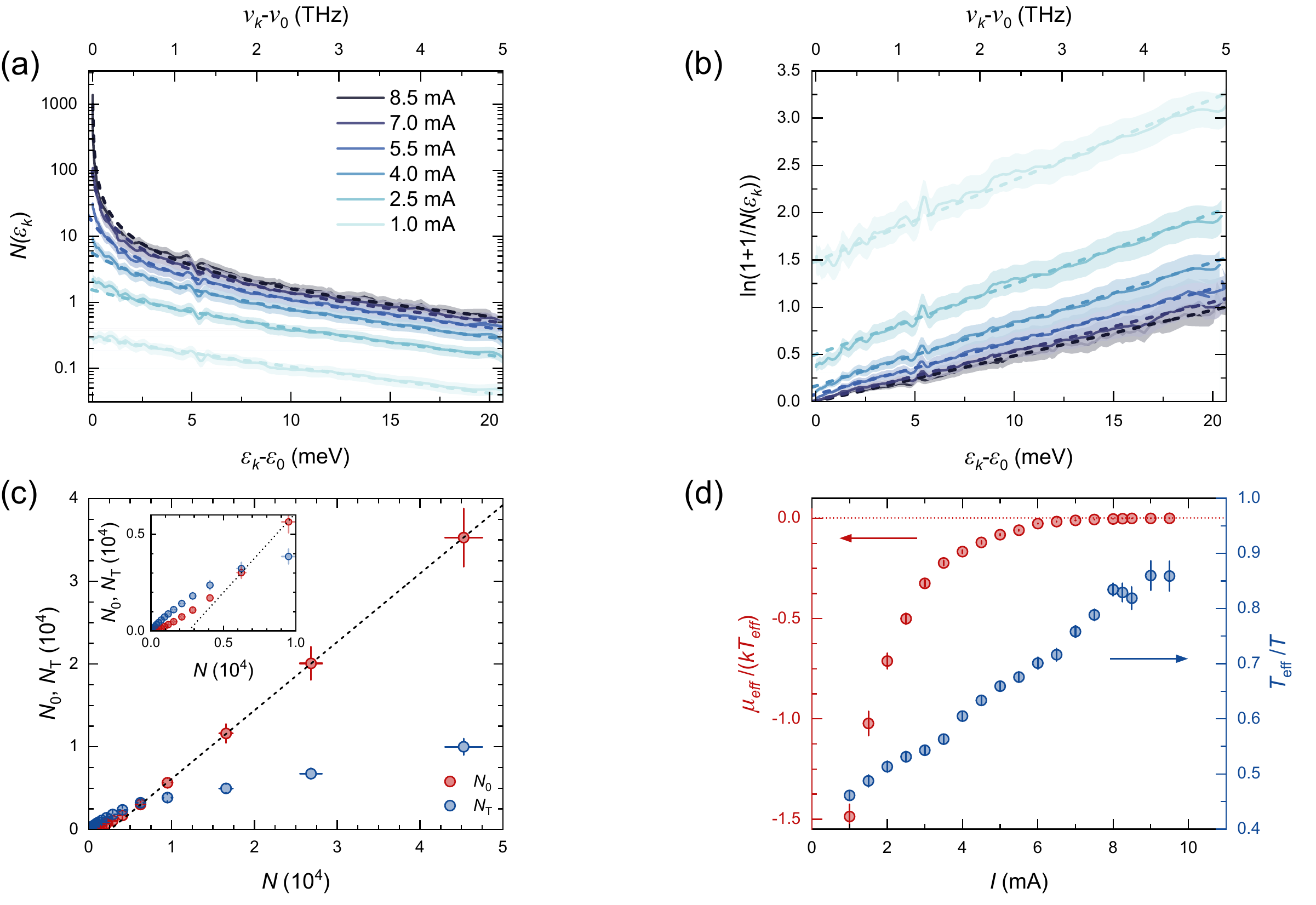}
\caption{\textbf{Experimental energy distributions and thermodynamic quantities} \textbf{(a)} Solid lines represent energy distributions extracted from the momentum spectra for different driving currents. \textbf{(b)} The same data is represented in logarithmic form (see text). In \textbf{(a)},\textbf{(b)} the energy scale is expressed with respect to the energy in the ground mode. The dashed lines are the fits of the BE distribution to the experimental data. The error bars, representing 95\% confidence intervals, are depicted as shaded regions. \textbf{(c)} Population of the ground state ($N_0$) and excited states ($N_T$) extracted from the experimental spectra. The dashed line is the linear fit above the condensation threshold to calculate the critical density ($N_C$). The inset shows the zoom-in into the low-number region of the main plot. \textbf{(d)} Thermodynamic quantities, effective chemical potential $\mu_\text{eff}$ and temperature $T_\text{eff}$ extracted from fitting the experimental distributions, as a function of driving current. The temperature of the heat sink is $T=293\,\text{K}$. Error bars in \textbf{(c)},\textbf{(d)} represent the 95\% confindence intervals.} 
\label{fig:Fig3}
\end{figure*}

To this end, we performed an analysis of the VCSEL spectral features, especially the distribution of occupations in the respective energy states. The investigated devices have large electrical apertures, resulting in a quasi-continuum of transversal optical modes (or in-plane energy states). Thus, photons in the resonator can be described by a parabolic dispersion in the in-plane direction $\varepsilon_k = \varepsilon_0 + \frac{\hbar^2 k^2}{2m_{\rm ph}}$ with an effective mass $m_{\rm ph}$. In our device $m_{\rm ph}\approx 2.75 \cdot 10^{-5}\,m_e$, where $m_e$ is the mass of the free electron. 

We employed the back focal plane (Fourier space or far-field) imaging technique to directly access the momentum dispersion, as shown schematically in Fig.~\ref{fig:Fig2}(a). The image is directed onto the monochromator slit, allowing for spectral analysis of the momentum dispersions. Momentum dispersion below $I<I_{\rm th}$ is presented in Fig.~\ref{fig:Fig2}(b). It shows thermalized distribution in momentum space, following the expected parabolic dispersion. The most distinguishing feature is observed above the threshold $I>I_{\rm th}$ (see Fig.~\ref{fig:Fig2}(c)), where the fundamental mode at $k_\parallel = 0$ dominates the spectrum. This is unusual behavior for such a large aperture resonator, as lasing in higher-order modes is commonplace~\cite{Degen1999}. We obtain this standard behavior in our lasing device with negative detuning, where right above the threshold current lasing in a higher-order mode is detected, together with a distinctive splitting in the momentum space (see Fig.~\ref{fig:Fig2}(d)). This crucial difference between the BEC and the lasing devices is confirmed in the spatially resolved spectra (near field), since in the case of BEC behavior we are dealing with a spatially homogeneous gas of photons, presented in Figs.~\ref{fig:Fig2}(e),(f), where condensation occurs in the fundamental transversal optical mode (ground state) of the system. In contrast, the lasing device operates in the higher-order mode, which is distributed closer to the aperture perimeter where the current density and optical gain are higher (see Fig.~\ref{fig:Fig2}(g))~\cite{Michalzik1998, Degen1999, Ackemann2000}.

We further explore the thermodynamic properties of the photon gas in the BEC device, by extracting the occupancies of the respective transversal energy states. Hence, we integrate the momentum-space electroluminescence data taking into account the density of states, the estimated photon lifetimes, and the efficiency of the optical setup (see Methods for details). The experimental energy distributions at different driving currents are presented in Fig.~\ref{fig:Fig3}(a). All data were successfully fitted with the BE distributions of Eq.~(\ref{eq:eq1}) by assuming a negligible $\Gamma$. Additional verification of the BEC distribution was also carried out, representing the data in logarithmic form, by transforming Eq.~(\ref{eq:eq1}) as $\ln[1+1/N(\varepsilon)] = \varepsilon/(k_B T)-\mu/(k_B T)$, which results as a linear function of energy (see Fig.~\ref{fig:Fig3}(b)). 

The data resemble the textbook behavior of a Bose condensed gas, such as massive occupation and  threshold-like dependency of the ground-state occupancy $N_0$ as a function of the total number of particles, along with saturation of the excited states $N_T$. These effects can be seen in the distributions in Fig.~\ref{fig:Fig3}(a). Figure~\ref{fig:Fig3}(c) summarizes the corresponding values of $N_0$, $N_T$. However, the thermal tails do not have the same slopes, which is more evident in Fig.~\ref{fig:Fig3}(b). This implies that, although the photons seem to be equilibrated, the temperature of the gas is not equal at different driving currents. Therefore, we denote the fitting parameters of the BE distribution as an effective chemical potential $\mu_\text{eff}$ and temperature $T_\text{eff}$, because these may not be equal to those set by the device conditions. 
Importantly, the geometry of the device imposes inhomogeneous current density across the aperture. Therefore, the chemical potential set by the quasi-Fermi levels and the temperature slightly vary spatially. The thermodynamic properties of the photon gas are a result of the spatially averaged overlap of the optical modes with the inhomogeneous QWs active medium~\cite{loirettepelous2023}. The results of the fits to the experimental data are presented in Fig.~\ref{fig:Fig3}(d). The effective chemical potential is always negative with respect to the fundamental mode energy and approaches zero when condensation occurs, supporting BEC behavior for an ideal gas. On the other hand, the effective temperature is a monotonic function of the driving current and saturates above the condensation transition to $T_\text{eff} \approx 250\,\text{K}$, which is approximately $T_\text{eff}/T \approx 0.85$ compared to the temperature of the heat sink $T = 293\,\text{K}$. Note that the actual temperature of the active region is expected to be even slightly higher due to the heating effects in the device (see Supplementary Information).  

From the data in Fig.~\ref{fig:Fig3}(c), we experimentally extracted the critical condensation value of the particles $N_C^\text{exp}  = 2604 \pm 91$. This value is expected at the condensation temperature $T\approx 220\,\text{K}$, which is in line with the experimental value extracted from Fig.~\ref{fig:Fig3}(d) at the condensation threshold $T_{\rm eff}\approx 223\,\text{K}$. 
\begin{figure}[t]
\centering
\includegraphics[width=\columnwidth]{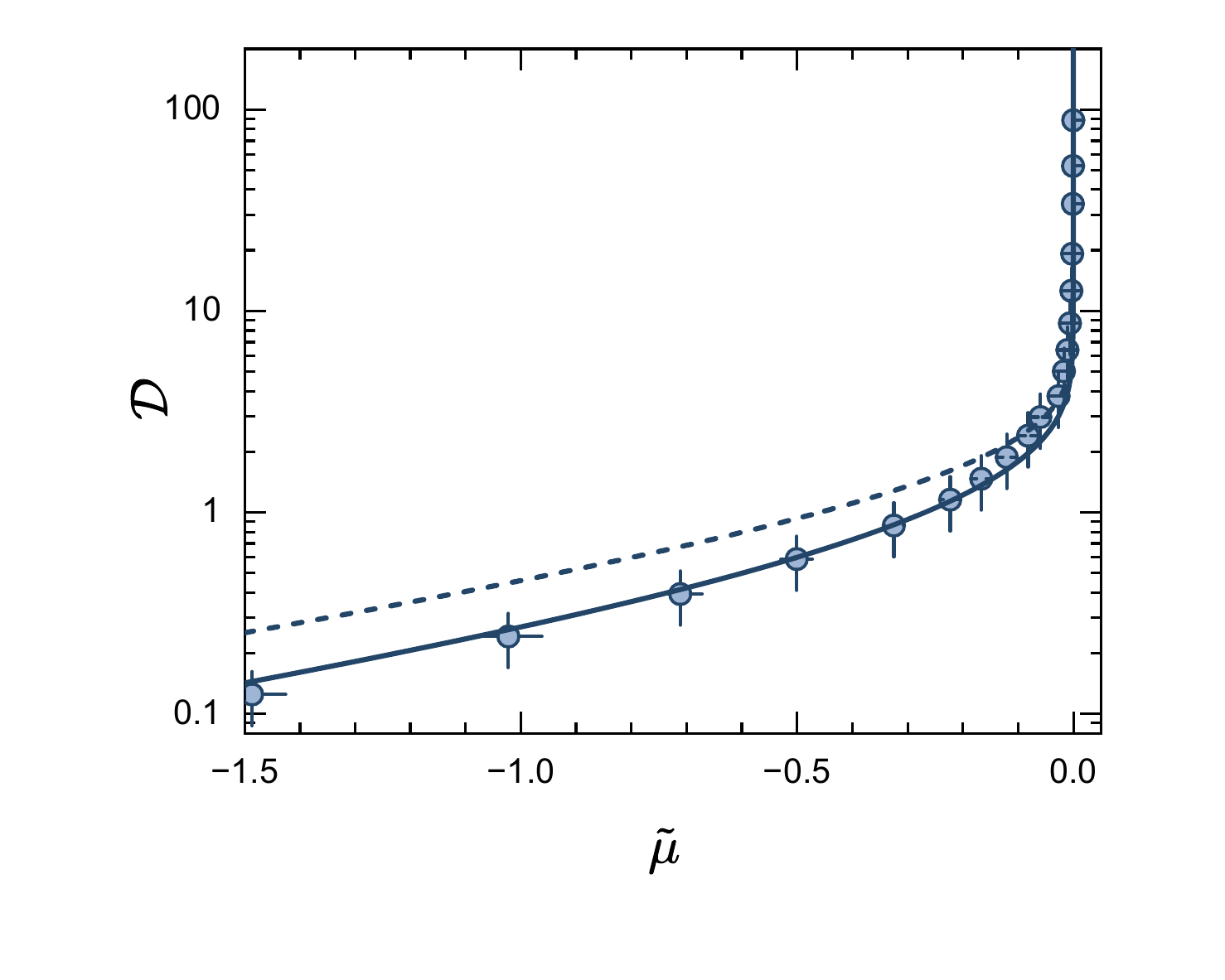}
\caption{\textbf{Determination of EOS} Points are extracted from the experimental data based on $T_\text{eff}$ and $\mu_\text{eff}$ (see main text). The dashed line is the theoretical EOS for a 2D Bose gas in the thermodynamic limit. The solid line is calculated by taking into account the finite collection angle of the optical setup. Error bars represent the 95\% confidence intervals.}
\label{fig:Fig4}
\end{figure}

All of these results suggest that we are dealing with a photonic gas that is not in full thermal and chemical equilibrium with the reservoir, which is the active region of the device. Equilibration to temperatures lower than the reservoir by stimulated cooling has recently been predicted for driven-dissipative bosonic condensates in the fast thermalization limit in a quantum model taking into account all correlations between states~\cite{Shishkov2022}. An experimental indication for the stimulated cooling effect can be seen in our data, as the occupations of the excited states are above unity in the condensed regime according to Fig.~\ref{fig:Fig3}(c) and there is a saturation of $T_\text{eff}$ above the condensation threshold in Fig.~\ref{fig:Fig3}(d). Therefore, it is interesting to examine what the EOS of the probed photon condensate is and whether it follows the EOS for a 2D Bose gas. The latter is written in the thermodynamic limit as
\begin{equation}
\mathcal{D}=- \ln \left[ 1+\exp \left( \frac{\mu}{k_B T}\right) \right]\,,
    \label{eq:eq2}
\end{equation}
where $\mathcal{D}=n \lambda_{T}^2$ represents the dimensionless phase space density. The photon density is defined by $n = N/(\pi R^2)$ with $\pi R^2$ denoting the surface area of the aperture and $R$ being its radius, while the thermal de Broglie wavelength of photons reads $\lambda_T= \sqrt{(2\pi\hbar^2)/(m_{\text{ph}} k_B T)}$. The EOS is expressed in normalized quantities by $\mathcal{D}$ and $\tilde{\mu} = \mu/(k_B T)$, hence the properties of the 2D bosonic gas are expected to be universal~\cite{Hung2011,Yefsah2011}.

The measured EOS, expressed by the experimental values $\mu_\text{eff}$ and $T_\text{eff}$, is presented in Fig.~\ref{fig:Fig4}. The data follow the equilibrium EOS, but with a larger slope in comparison to the thermodynamic limit. This can be explained by the finite collection angle of the collection optics in our setup, which is represented by the numerical aperture  (NA) of the microscope objective. We cannot detect energies emitted beyond the maximal angle. Numerical calculations confirm the observations, as we computed the phase-space density for a finite number of states defined by the NA. The results of the calculations presented in Fig.~\ref{fig:Fig4} as a solid line perfectly match the experimental data. The discrepancy of the experimental data from the EOS in the thermodynamic limit is analogous to previous reports~\cite{Busley2022}, where the finite trap depth was given to explain the lower than expected phase-space density. In our case, the energies of all possible transversal modes in the VCSEL are expected to go beyond the values dictated by the objective NA.

We have demonstrated that emission from a positively detuned VCSEL has the properties of a homogeneous 2D Bose-Einstein condensed gas of photons in a finite system. The measured nonequilibrium nature of the gas can be a signature of reaching the fast, stimulated thermalization limit, because the cavity is characterized by a relatively short photon lifetime. Photon condensation in semiconductor resonators offers the possibility of observing the superfluidity of a weakly interacting Bose gas. Photon interactions are expected to be mediated by semiconductor non-linearity, which is significantly enhanced by the cavity and has a subpicosecond-order response time~\cite{Kriso2021,Yuce2016}. There are no clear indications of such interactions in our data, because the cavity energy shifts are dominated by the current- and temperature-induced changes in the refractive index. Further studies are needed, focused on probing the hydrodynamics of the condensed photons directly, by perturbing them from the steady state~\cite{Estrecho2021, Peinke2021}. Nevertheless, in addition to material non-linearities, the dissipative nature of the photon gas encourages further studies of phase ordering~\cite{Gladilin2020} and universal scaling in a 2D geometry~\cite{Fontaine2022,Bloch2022} and signs of non-Hermitian effects~\cite{Khurgin2020}. 

Another direction for future work is to test the fluctuations of the non-equilibrium BEC and to compare it to the BEC in thermal equilibrium~\cite{Christensen2021,Ozturk2023} as well as to standard VCSEL operation~\cite{Agarwal1972,Assmann2009}. The mature technology of semiconductor VCSELs offers the possibility of utilizing the BEC regime to achieve single-mode emission from large aperture devices characterized by excellent beam quality, without the need for sophisticated additional fabrication and processing of the laser mesa~\cite{Contractor2022,Yang2022,Yoshida2023}. BEC VCSELs could also be applied in more complex lattice geometries, to study topological effects in well-controlled current-operated devices at room temperature~\cite{Dikopoltsev2021}.
\section*{Methods}
\subsection*{Thermalization of photons in a semiconductor laser}
The principles of light absorption and recombination in an excited semiconductor QW, depicted in Fig.~\ref{fig:Fig1}(b), can be described by the following transition rates~\cite{loirettepelous2023,Coldren2012} for emission
\begin{equation}
    R_{\rm em}(\varepsilon) = R(\varepsilon)f_c(\varepsilon,T,\mu_c)\big[1-f_v(\varepsilon,T,\mu_v)\big]
\end{equation}
and absorption
\begin{equation}
    R_{\rm abs}(\varepsilon) = R(\varepsilon)f_v(\varepsilon,T,\mu_v)\big[1-f_c(\varepsilon,T,\mu_c)\big]\,,
\end{equation}
where $f_{c,v}=\left\{\exp\left[ ( \varepsilon-\mu_{c,v})/(k_{B}T) \right]+1\right\}^{-1}$ denote the thermalized Fermi-Dirac distributions of electrons in the conduction and holes in the valence bands, respectively. $R(\varepsilon)$ stands for the transition rate at energy $\varepsilon$, taking into account the photonic and electronic density of states, the overlap of the optical modes with the active medium, and the intrinsic properties of the active medium itself~\cite{loirettepelous2023}. The natural consequence in semiconductors is the van Roosbroeck-Schockley relation, which appears, after some algebra, from the relation
\begin{equation}
\label{RSR}
\frac{R_{\rm abs}(\varepsilon)}{R_{\rm em}(\varepsilon)}=\exp \left(\frac{\varepsilon-\mu}{k_{B}T}\right)
\end{equation}
with $\mu = \mu_c-\mu_v$~\cite{Wurfel1982,loirettepelous2023,Coldren2012}.

Now, the rate equation for the occupation of an optical mode at $\varepsilon$ is expressed as
\begin{equation}
    \frac{d}{dt} N(\varepsilon) = R_\text{em}(\varepsilon) \big[N(\varepsilon)+1\big] - \big[R_\text{abs}(\varepsilon) + \gamma(\varepsilon)\big]N(\varepsilon)
\end{equation}
where $\gamma(\varepsilon)=1/\tau(\varepsilon)$ denotes the decay rate of a photon from an empty cavity at $\varepsilon$.
Thus, the resulting steady-state solution gives
\begin{equation}
    N(\varepsilon) = \frac{R_\text{em}(\varepsilon) }{\gamma(\varepsilon)-\left[R_\text{em}(\varepsilon)-R_\text{abs}(\varepsilon)\right]}\, .
\end{equation}
After dividing both nominator and denominator by $R_\text{em}(\varepsilon)$ as well as using the van Roosbroeck-Schockley relation (\ref{RSR}) we obtain for $N(\varepsilon)$ the result of Equation~(\ref{eq:eq1}). This
amounts to a Bose-Einstein distribution with the correction parameter $\Gamma(\varepsilon)=\gamma(\varepsilon)/R_{\rm em}(\varepsilon)$. 

We estimated this correction parameter $\Gamma(\varepsilon_0)$ for the fundamental mode $\varepsilon_0$ of the device as follows. The decay rate of a photon from an empty cavity follows from the decay time calculated from the realistic numerical model: $\gamma(\varepsilon_0) = 1/\tau (\varepsilon_0) = 1/\left(3.04~\text{ps}^{-1}\right) \approx 0.33~\text{ps}^{-1}$ (see Supplementary information). We are able to determine the value of $R_\text{em}(\varepsilon_0)=42 \pm 3~\text{ps}^{-1}$ close to the threshold by measuring the linewidth dependence of the ground mode as a function of occupation below the condensation threshold \cite{Perez2014}. With this, we obtain the value $\Gamma(\varepsilon_0) \approx 0.008$ as mentioned above.
\subsection*{Sample} 
The VCSEL epitaxial structure is designed for high-speed data communication at 980\,nm. The
epitaxial structure is monolithically grown on an n-doped GaAs substrate. The multi-quantum well (MQW) active region is composed of 5 $\rm In_{0.23}Ga_{0.77}As$ QWs and
6 $\rm GaAs_{0.86}P_{0.14}$ barriers centered in $\mathrm {Al}_{x}\mathrm{Ga}_{1-x}\mathrm{As}$ cavity graded from $x = 0.38$ to $0.80$ with an optical cavity thickness of $\lambda/2$. The cavity is sandwiched by 15.5-pair
GaAs/$\rm Al_{0.9}Ga_{0.1}As$ top and 37-pair bottom distributed Bragg reflector (DBR) mirrors. The top and
bottom DBRs are C-doped for the p-type and Si-doped for the n-type, respectively. In both mirrors,
graded interfaces are incorporated for lower electrical resistance of the structure. Importantly, two 20\,nm thick
$\rm Al_{0.98}Ga_{0.02}As$ layers are placed to form oxide apertures in the first nodes of the standing
wave at the top and bottom of the cavity. These oxide layers are halfway in the optical cavity and halfway in the first pair of layers in the DBRs.

The VCSELs are processed using standard top-down photolithography. In the first step, the
Ti/Pt/Au p-type contact rings are deposited with the use of electron beam deposition (E-beam). The
mesa structures are then patterned and etched using inductively coupled plasma reactive-ion etching
(ICP-RIE) in a $\rm Cl_2 /BCl_3$-based plasma. After etching, current confinement apertures are formed
by selective wet thermal oxidation of $\rm Al_{0.98}Ga_{0.02}As$ layers in an oxidation oven in a nitrogen atmosphere
with overpressure of water vapor and at high temperature (420\textdegree C). In the following step, horseshoe-shaped Ni/AuGe/Au n-type contact pads are deposited and annealed in a rapid thermal processing
(RTP) furnace. The structures are then planarized with the use of a spin-on dielectric polymer of benzocyclobutene (BCB). The BCB layer is patterned with the use of photolithography and RIE etching in a $\rm CF_4$-based plasma
to selectively open surface areas to the bias to the p- and n-type contacts. In the final step, the ground-signal-ground (GSG)
Cr/Pt/Au contact pads are deposited.
\subsection*{Experimental setup}
The sample used in this study was a fully processed quarter of the whole epitaxial wafer. The sample was placed on a thermo-electrically cooled plate (Thorlabs PTC1) with an additional temperature sensor buried inside a custom heatsink plate mounted on the top, to ensure control of the temperature in the close vicinity of the sample. The temperature of the heatsink was set to 20\,\textdegree C throughout all experiments. The temperature-controlled plate was placed on a manual translation stage. The sample was contacted by a microwave probe (GGB Picoprobe 40A) located in an additional manual translation stage. The devices were driven with a direct current by a stabilized precise source/measure unit (Keysight B2901B). 

The device emission was collected using an infinity-corrected objective of NA = 0.65 (Mitutoyo M Plan Apo NIR HR 50x). As described in the main text, to measure the momentum spectra (far field) we imaged the back focal plane of the objective with a set of achromatic lenses onto the 0.3\,m-focal length monochromator entrance slit (Princeton Instruments Acton SP-2300i) and the electroluminescence signal was dispersed through a grating (1200 grooves/mm) onto an electron-multiplied charge-coupled device (EMCCD -- Teledyne Princeton Instruments ProEM-HS:1024BX3). To record the spatially resolved spectra (near field), one of the lenses was removed from the optical path, which enabled projection of the real-space image onto the monochromator slit. This lens was mounted on a flip mount, allowing quick and convenient switching between the two measurement modes of the setup.
\subsection*{Analysis of the momentum space}
Taking advantage of homogeneous emission from the BEC device, we determined the thermodynamic properties of the photon gas from the momentum space. We extracted the mean photon occupation distribution by integrating the momentum space emission, using the standard procedure used in cavity-polariton physics~\cite{Kasprzak2006,Pieczarka2019}.

The mean number of photons collected at a CCD pixel row representing a chosen $k$-state is represented as follows:
\begin{equation}
    N_{\rm ph}(k) = \eta \frac{dN_{\rm CCD}(k)}{dt} \tau(k),
\end{equation}
where $\eta$ is the calibrated collection efficiency of our setup, $dN_{\rm CCD}(k)/dt$ is the count rate per second on the CCD camera pixel, and $\tau(k)$ is the photon lifetime at $k$. The photon lifetime was estimated from the experiment by extracting the emission linewidth $\Delta \varepsilon_k = \hbar/\tau(k)$~\cite{Pollnau2020} by fitting a Lorentzian function to the data from a $k$-state pixel row.

Subsequently, the occupation number at the $k$-state is calculated taking into account the number of states subtended by a pixel at $k$-position in cylindrical coordinates $N_{\rm st}(k)=k \Delta k \Delta \phi (4\pi / S)^{-1}$, where $S$ is the surface area of the device aperture. The number of states in momentum space was confirmed by numerical simulations of the optical modes confined in the device (see Supplementary Information). The final expression is the following:
\begin{equation}
    N(\varepsilon (k))=\frac{N_{\rm ph}(k)}{N_{\rm st}(k)}= \frac{4 \pi^2 \eta}{2 k \Delta k \Delta \phi S }\frac{dN_{\rm CCD}(k)}{dt} \tau(k),
\end{equation}
which also considers the spin degeneracy 2 of all states, as our experiment was not polarization resolved. The energy $\varepsilon(k)$ of the measured $k$-state is extracted from the fitted Lorentzian peak.

\subsection*{Acknowledgments}
We gratefully thank Maciej Dems for his support in improving the numerical simulation codes used in this work and Milan Radonji\'{c} for valuable discussions. MP and ANP acknowledge support from the Polish National Science Center, grant Sonata no. 2020/39/D/ST3/03546. TC acknowledges the project Sonata Bis no. 2015/18/E/ST7/00572 from the Polish National Science Center, within which the VCSELs used in this work were fabricated. AP acknowledges financial support by the Deutsche Forschungsgemeinschaft (DFG, German Research Foundation) via the Collaborative Research Center SFB/TR185 (Project No. 277625399).

\subsection*{Author contributions}MP conceived this research project. MP and ANP conducted the experiments, and MP performed the detailed data analysis. JAL designed the epitaxial structure and provided the planar wafer sample. MG designed the laser mesa outline and performed all fabrication steps. TC performed the numerical modeling of the devices. MP, AP, MW, and TC contributed to the theoretical analysis and interpretation of the data. All authors discussed the results. MP wrote the manuscript with input from all authors.

\bibliography{references}

\end{document}